\documentclass[aps,prd,twocolumn,amsmath,amssymb,showpacs]{revtex4}

\usepackage{graphicx}
\usepackage{dcolumn}
\usepackage{bm}

\begin{document}

\title{Complex angular momentum in black hole physics and quasinormal
modes}

\author{Yves D\'ecanini}
\email{decanini@univ-corse.fr}
\author{Antoine Folacci}
\email{folacci@univ-corse.fr} \affiliation{
UMR CNRS 6134 SPE, Equipe Physique Semi-Classique (et) de la Mati\`ere Condens\'ee \\
Universit\'e de Corse, Facult\'e des Sciences, BP 52, 20250 Corte,
France}
\author{Bruce Jensen}
\email{Bruce.Jensen@marconi.com} \affiliation{Department of
Mathematics, University of Southampton, Southampton SO17 1BJ,
United
Kingdom  \\
and Marconi Communications, Brindle Avenue, New Century Park,
Coventry CV3 1HJ, United Kingdom}

\date{\today}

\begin{abstract}

By using the complex angular momentum approach, we prove that the
quasinormal mode complex frequencies of the Schwarzschild black
hole are Breit-Wigner type resonances generated by a family of
surface waves propagating close to the unstable circular photon
(graviton) orbit at $r=3M$. Furthermore, because each surface wave
is associated with a given Regge pole of the $S$-matrix, we can
construct semiclassically the spectrum of the quasinormal-mode
complex frequencies from Regge trajectories. The notion of surface
wave orbiting around black holes thus appears as a fundamental
concept which could be profitably introduced in various areas of
black hole physics in connection with the complex angular momentum
approach.
\end{abstract}


\maketitle

\section{Introduction}

Since the pioneering work of Watson \cite{Watson18} dealing with
the propagation and diffraction of radio waves around the Earth,
the complex angular momentum (CAM) method has been extensively
used in several domains of scattering theory (see the monographs
of Newton \cite{New82} and of Nussenzveig \cite{Nus92} and
references therein for various applications in quantum mechanics,
nuclear physics, electromagnetism, optics, acoustics and
seismology). The success of the CAM method is due to its ability
to provide a clear description of a given scattering problem by
extracting the physical information (linked to the geometrical and
diffractive aspects of the scattering process) which is hidden in
partial-wave representations.

The CAM method was first used in gravitational physics by
Chandrasekar and Ferrari \cite{Chandra} in their study of
nonradial oscillations of relativistic stars. They used the theory
of Regge poles \cite{New82} to determine the flow of gravitational
energy through the star. The general framework for the CAM
description of Schwarzschild black hole scattering was developed
by Andersson and Thylwe \cite{Andersson1}, which Andersson
\cite{Andersson2} then used to interpret the black hole glory. An
important concept, which is naturally present in the CAM point of
view, is that of a surface wave, which allows a description of the
diffractive effects of scattering. Andersson established that the
surface waves orbiting around the Schwarzschild black hole of mass
$M$ propagate close to the unstable photon orbit at $r=3M$. Yet
except for these articles, the CAM/surface-wave approach to
resonant scattering in black hole physics has been neglected in
favor of the quasi-normal modes (QNMs), in which the dynamical
response to an external perturbation is explained in terms of
resonant frequencies. For recent reviews of the QNM approach see
Refs.~\onlinecite{Kokkotas} and~\onlinecite{Nollert} as well as
chapter 4 of Ref.~\onlinecite{Frolov-Novikov}. An introduction to
black hole scattering can be found in \cite{AndJens}.

In this paper we will establish the connection between Andersson's
surface waves and the QNMs. More precisely, we shall prove here
that the QNM complex frequencies of the Schwarzschild black hole
are Breit-Wigner-type resonances generated by the surface waves.
Moreover, because each surface wave is associated with a given
Regge pole of the $S$-matrix, we can construct the spectrum of the
QNM complex frequencies from the Regge trajectories, i.e., from
the curves traced out in the CAM plane by the Regge poles as a
function of the frequency.

As early as 1972, it was suggested by Goebel \cite{Goebel} that
the black hole normal modes could be interpreted in terms of
gravitational waves in spiral orbits close to the unstable photon
orbit at $r=3M$, which decay by radiating away energy. In the
present paper, using the CAM approach, we establish this appealing
and physically intuitive picture on a rigorous basis. To conclude,
we also provide a framework for future developments.

\section{From CAM to QNM}

We first consider the scattering of a monochromatic scalar wave,
with time dependence $\exp (-i\omega t)$, by the Schwarzschild
black hole of mass $M$. The corresponding scattering amplitude can
be written (see, e.g. \cite{Frolov-Novikov})
\begin{equation}\label{ampli}
f(\omega, \theta)=\frac{1}{2i\omega} \sum_{\ell=0}^{+\infty}
(2\ell + 1)\left(S_\ell(\omega) - 1 \right)P_{\ell}(\cos \theta)
\end{equation}
where $\ell$ is the ordinary angular momentum index, $P_\ell $ are
the usual Legendre polynomials and $S_\ell $ are the diagonal
elements of the $S$-matrix. For a given angular momentum $\ell$,
the coefficient $S_\ell $ is obtained from the partial wave
solution $\Phi_\ell $ of the following problem:

\noindent (i) $\Phi_\ell $ satisfies the Schr{\"o}dinger-type
equation
  \begin{equation}\label{RW}
  \frac{d^2 \Phi_\ell}{dr_*^2} + \left[ \omega^2 - V(r)\right]
  \Phi_\ell=0.
  \end{equation}
Eq.~(\ref{RW}) is obtained from the scalar wave equation after
separating variables and is called the Regge-Wheeler equation.
Here $r$ is the standard radial Schwarzschild coordinate and
$r_*=r+2M\ln \left(1-2M/r \right) + \mathrm{const}$ is the
Regge-Wheeler tortoise coordinate while the potential $V(r)$ is
given by
\begin{equation}\label{scpot}
V(r) = \left(\frac{r-2M}{r} \right) \left[
\frac{\ell(\ell+1)}{r^2} +\frac{2M}{r^3}\right].
\end{equation}

\noindent (ii) $\Phi_\ell $, as any physical wave, must have a
purely ingoing behavior at the event horizon at $r=2M$, i.e.
\begin{equation}
\label{bc1} \Phi_\ell (r) \underset{r_* \to -\infty}{\sim}
e^{-i\omega r_* }.
\end{equation}

\noindent (iii) At spatial infinity $r \to +\infty$, $\Phi_\ell $
has the asymptotic behavior
\begin{equation}
\label{bc2} \Phi_\ell(r) \underset{r_* \to +\infty}{\sim}
\frac{1}{T_\ell(\omega)}e^{-i\omega r_* + i\ell\pi/2}-
\frac{S_\ell(\omega)}{T_\ell(\omega)} e^{+i\omega r_* -
i\ell\pi/2}.
\end{equation}

For certain complex values of $\omega $, both $S_\ell$ and
$T_\ell$ have a simple pole but $S_\ell/T_\ell$ is regular. These
values are the frequencies of the QNMs, which we can define (see
Eqs.~(\ref{bc1}) and (\ref{bc2})) as the solutions of the wave
equation which represent a purely outgoing wave at infinity and a
purely ingoing wave at the horizon. We now denote by $\omega_{\ell
n}=\omega^{(o)}_{\ell n}-i\Gamma _{\ell n}/2$ with
$\omega^{(o)}_{\ell n}>0$ and $\Gamma _{\ell n}>0$ the QNM
frequencies, $\omega^{(o)}_{\ell n}$ representing the frequency of
the oscillation corresponding to the QNM and $\Gamma _{\ell n}$
representing its damping. In the immediate neighborhood of
$\omega_{\ell n}$, $S_\ell(\omega )$ has the Breit-Wigner form,
i.e.,
\begin{equation}\label{BW}
S_\ell(\omega ) \propto \frac{\Gamma _{\ell p}/2}{\omega
-\omega^{(o)} _{\ell p}+i\Gamma _{\ell p}/2}.
\end{equation}

\begin{figure}
\includegraphics[height=5.0cm,width=8cm]{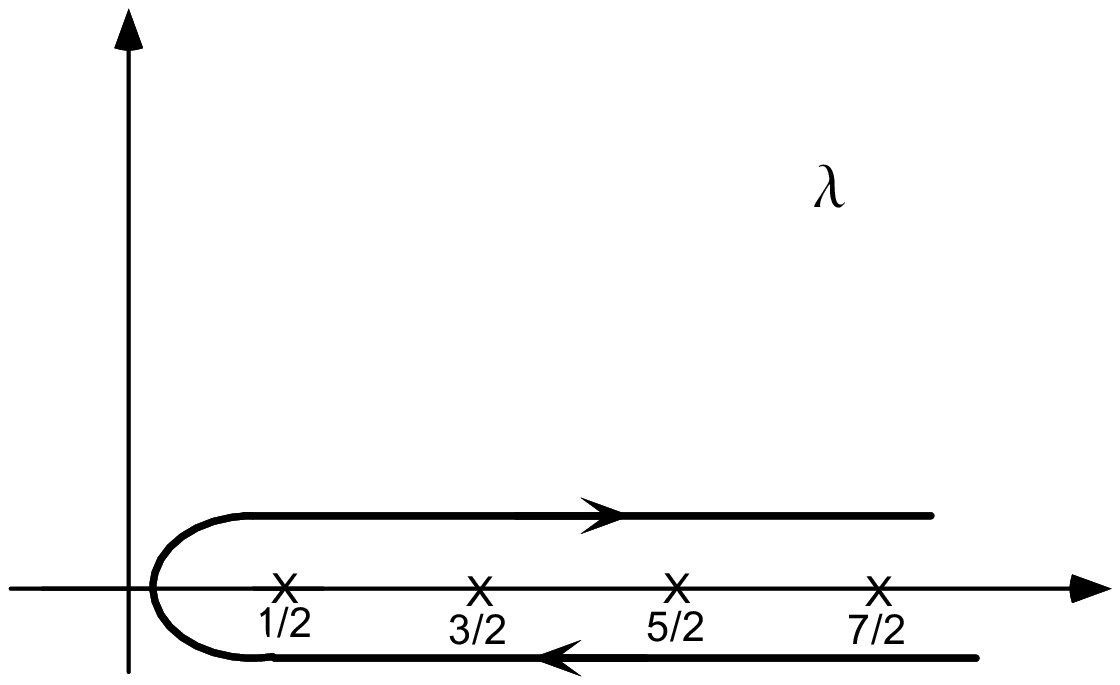}
\caption{\label{fig:watson} The Watson integration contour. }
\end{figure}

Using the CAM method, we can provide a physical picture of the
scattering process in term of diffraction by surface waves and a
physical explanation of the mechanism of QNM excitation valid for
high frequencies. By means of a Watson transformation
\cite{Watson18} applied to the scattering amplitude (\ref{ampli}),
we can write \cite{Andersson1}
\begin{equation}\label{ampliII}
f(\omega, \theta)=-\frac{1}{2\omega} \int_{\cal C} \frac{\lambda
\left(S_{\lambda -1/2 } (\omega) - 1 \right)}{\cos \pi
\lambda}P_{\lambda -1/2}(-\cos \theta) ~d\lambda .
\end{equation}
Here $\mathcal{C}$ is the integration contour in the complex
$\lambda$-plane \cite{Watson18} illustrated in
Fig.~\ref{fig:watson}. The Watson transformation permits us to
replace the ordinary angular momentum $\ell $ by the complex
angular momentum (CAM) $\lambda$. Here $S_{\lambda -1/2} (\omega)$
is now an analytic extension of $S_\ell (\omega )$ into the
complex $\lambda$-plane which is regular in the vicinity of the
positive real $\lambda$ axis. Moreover, $P_{\lambda-1/2} (z) $ is
the hypergeometric function $F(-\lambda +1/2 , \lambda +1/2; 1;
(1-z)/2)$.

We can then deform the path of integration in Eq.~(\ref{ampliII}),
taking into account the possible singularities. The only
singularities that are encountered are the poles of the S-matrix
lying in the first quadrant of the CAM plane \cite{Andersson1}.
They are known as Regge poles \cite{New82,Nus92} and we shall
denote them by $\lambda_n (\omega)$, the index $n=1,2, \dots $
permitting us to distinguish between the different poles. By
Cauchy's Theorem we can then extract from Eq.~(\ref{ampliII}) the
contribution of a residue series over Regge poles given by
\begin{equation}\label{ampliPR1}
f_P(\omega, \theta)=\frac{-i\pi}{\omega} \sum_{n=1}^{+\infty}
\frac{\lambda_n(\omega )r_n(\omega)}{\cos \left(\pi
\lambda_n(\omega )  \right)} P_{\lambda_n(\omega )-1/2 }(-\cos
\theta)
\end{equation}
where $r_n(\omega )=\mathrm{residue}\left(S_{\lambda-1/2}(\omega
)\right)_{\lambda = \lambda _n(\omega )}$. It should be noted that
$f$ differs from $f_P$ by a background integral \cite{Andersson1}
which does not play any role in the resonance phenomenon. By using
the asymptotic expansion
\begin{equation}
P_{\lambda -1/2}(-\cos \theta) \sim \frac{e^{i\lambda (\pi -
\theta)-i\pi/4}+e^{-i\lambda (\pi - \theta)+i\pi/4}}{({2\pi\lambda
\sin \theta})^{1/2}} \nonumber
\end{equation}
as $|\lambda| \to \infty$, valid for $|\lambda| \sin \theta >1$,
as well as
\begin{equation}
\frac{1}{\cos \pi \lambda }=2i\sum_{m=0}^{+\infty} e^{ i\pi(2m+1)(
\lambda-1/2)} \nonumber
\end{equation}
which is true if $\mathrm{Im} \ \lambda
> 0$, we can write
\begin{eqnarray}\label{ampliPR2}
&   &  f_P(\omega, \theta) = \frac{2\pi}{i\omega}
\sum_{n=1}^{+\infty} \frac{\lambda_n(\omega )r_n(\omega)}{\left[
2\pi \lambda_n(\omega
) \sin \theta \right]^{1/2}}  \nonumber \\
&  & \qquad  \qquad  \times \sum_{m=0}^{+\infty}  \left(
e^{i\lambda_n(\omega
)(\theta +2m\pi) -im\pi +i\pi/4} \right.   \nonumber \\
&  & \qquad \qquad \qquad \qquad  + \left. e^{i\lambda_n(\omega
)(2\pi - \theta +2m\pi) -im\pi - i\pi/4} \right). \quad
\end{eqnarray}
In Eq.~(\ref{ampliPR2}), terms like $\exp[i\lambda_n(\omega
)(\theta)]$ and $\exp[i\lambda_n(\omega )(2\pi - \theta)]$
correspond to surface wave contributions. Because a given Regge
pole $\lambda_n(\omega )$ lies in the first quadrant of the CAM
plane, $\exp[i\lambda_n(\omega )(\theta)]$ (resp.
$\exp[i\lambda_n(\omega )(2\pi - \theta)]$) corresponds to a
surface wave propagating counterclockwise (resp. clockwise) around
the black hole and $\mathrm{Re} \ \lambda_n(\omega)$ represents
its azimuthal propagation constant while $\mathrm{Im} \
\lambda_n(\omega)$ is its damping constant. The exponential decay
$\exp[-\mathrm{Im} \ \lambda_n(\omega)\theta]$ (resp.
$\exp[-\mathrm{Im} \ \lambda_n(\omega)(2\pi - \theta)]$) is due to
continual reradiation of energy. Moreover, in
Eq.~(\ref{ampliPR2}), the sum over $m$ takes into account the
multiple circumnavigations of the surface waves around the black
hole as well as the associated radiation damping. Finally, the
presence of the factor $\exp[-im\pi]$ in Eq.~(\ref{ampliPR2})
should be also noted: it accounts for the phase advance due to the
two caustics on the scattering axis.

The resonant behavior of the black hole can now be understood in
terms of surface waves. As $\omega $ varies, each Regge pole
$\lambda_n(\omega )$ describes a Regge trajectory \cite{New82} in
the CAM plane. When the quantity $\mathrm{Re} \ \lambda_n(\omega
)$ coincides with a half-integer (a half-integer but not an
integer because of the caustics), a resonance occurs. Indeed, it
is produced by a constructive interference between the different
components of the $n$-th surface wave, each component
corresponding to a different number of circumnavigations.
Resonance wave frequencies $\omega ^{(o)}_{\ell n}$ are therefore
obtained from the Bohr-Sommerfeld type quantization condition
\begin{equation}\label{sc1}
\mathrm{Re} \ \lambda_n \left(\omega ^{(o)}_{\ell n}  \right)=
\ell + \frac{1}{2} \qquad \ell =0,1,2,\dots .
\end{equation}
By assuming that $\omega $ is in the neighborhood of
$\omega^{(o)}_{\ell n}$ and using $\mathrm{Re} \ \lambda_n (\omega
) \gg \mathrm{Im} \ \lambda_n (\omega )$ (which can be numerically
verified, except for very low frequencies), we can expand
$\lambda_n(\omega)$ in a Taylor series about $\omega^{(o)}_{\ell
n}$, and obtain
\begin{eqnarray} \label{TS}
&&\lambda_n(\omega) \approx \ell + \frac{1}{2} +  \left. \frac{d \
\mathrm{Re} \lambda_n(\omega)}{d\omega} \right|_{\omega =\omega
^{(o)}_{\ell n}} (\omega - \omega ^{(o)}_{\ell n}  )    \nonumber
\\
 && \qquad \qquad + i \ \mathrm{Im} \ \lambda_n (\omega
^{(o)}_{\ell n}).
\end{eqnarray}
Then, by replacing Eq.~(\ref{TS}) in the term ${\cos \left(\pi
\lambda_n(\omega )  \right)}$ of Eq.~(\ref{ampliPR1}), we show
that $f_P(\omega, \theta)$ presents a resonant behavior given by
the Breit-Wigner formula (\ref{BW}) with
\begin{equation}\label{sc2}
\frac{\Gamma _{\ell n}}{2}= \left.  \frac{\mathrm{Im} \ \lambda_n
(\omega )}{d \ \mathrm{Re} \ \lambda_n (\omega ) /d\omega }
\right|_{\omega =\omega ^{(o)}_{\ell n}}.
\end{equation}
Eqs.~(\ref{sc1}) and (\ref{sc2}) are kind of semiclassical
formulas which permit us to determine the location of the
resonances from Regge trajectories.

We now discuss the numerical aspects of our work. In order to
determine the location of the QNM frequencies from
Eqs.~(\ref{sc1}) and (\ref{sc2}), we need the Regge trajectories.
For a given $\omega$ real and positive, we search for a complex
value of $\ell$ such that the coefficient $1/T_\ell(\omega)$ is
zero but $S_\ell(\omega)/T_\ell(\omega)$ is not. Such a value is
then a pole of $S_\ell(\omega)$ and provides immediately the value
of the associated pole for $S_{\lambda -1/2}(\omega)$. Leaver
\cite{LeaverI} presents a method for finding the zeros of
$1/T_\ell(\omega)$. Though he uses his method to find the QNM
frequencies, his method is equally valid, mutatis mutandis, for
finding the  Regge poles $\{\lambda_n=\ell_n +1/2 \}$ of
$S_\ell(\omega)$.
 Leaver writes the
solution to Eq.~(\ref{RW}) as an infinite sum
\begin{eqnarray}
&&\Phi_\ell(r) = (r-2M)^\rho \left(\frac{2M}{r}\right)^{2\rho}
e^{-\rho \left( \frac{r-2M}{2M} \right)} \nonumber \\
&& \qquad \qquad \qquad  \times \sum_{k=0}^\infty a_k
\left(\frac{r-2M}{r}\right)^k
\end{eqnarray}
where $\rho= -i2M\omega$. The recurrence relations between the
$a_k$ are given by Leaver:
\begin{equation}
\alpha_k a_{k+1} + \beta_k a_k + \gamma_k a_{k-1} = 0, \label{rec}
\end{equation}
where
\begin{eqnarray}
\alpha_k &=& k^2 + 2k (\rho+1) + 2\rho+1,    \\
\beta_k &=& -[2k^2 + 2k(4\rho+1)+8\rho^2 +4\rho+\ell(\ell+1)+1],
\nonumber  \\
   \\
\gamma_k &=& k^2 +4 k\rho + 4\rho^2.
\end{eqnarray}
Leaver noted that the coefficient $1/T_\ell(\omega)$ has a zero
whenever the sum $\sum a_k$ converges.
\begin{figure}
\includegraphics[height=6cm,width=8cm]{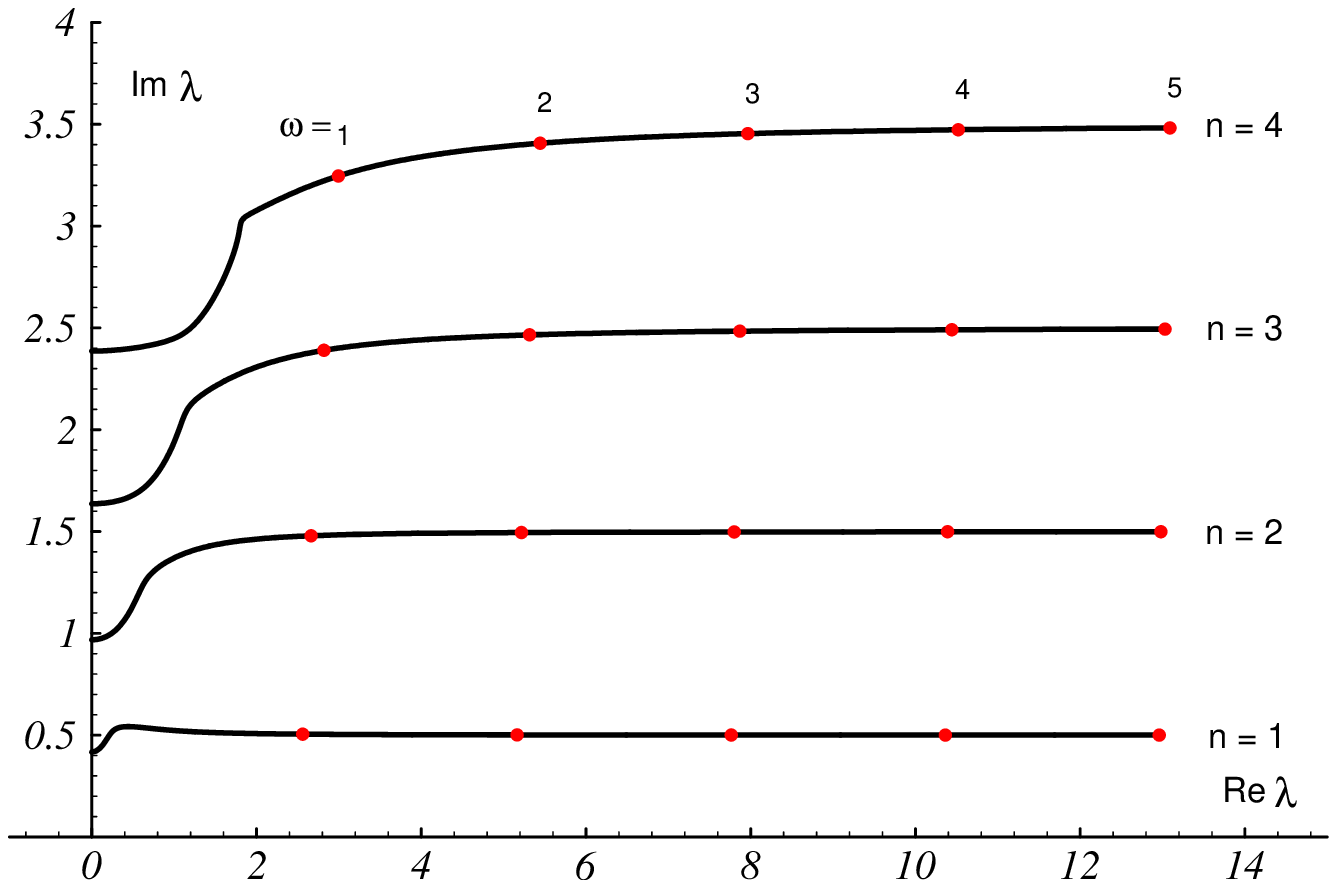}
  \caption{\label{fig:suivi-s0} The Regge poles $\lambda_n(\omega)$ followed
for $\omega=0\to 5$, $n=1,2,3,4$ for a scalar perturbation
($2M=1$).}
\end{figure}
\begin{table}
\caption{\label{tab:table1} A sample of QNM frequencies for scalar
perturbations ($2M=1$).}
\begin{ruledtabular}
\begin{tabular}{cccccc}
&       &\quad Exact\footnotemark[1] \quad& \quad
Exact\footnotemark[1]\quad&  Semiclassical  &  Semiclassical  \\
 $\ell $ & $n$ &$   \omega ^{(o)}_{\ell n}  $
 &  $  \Gamma _{\ell n}/2  $
 & $  \omega ^{(o)}_{\ell n}   $ &  $ \Gamma _{\ell n}/2  $  \\
\hline
0& 1 & 0.22091   & -0.20979  & 0.19009  & -0.23120  \\
 & 2 & 0.17223   & -0.69610  & 0.09743  & -0.33189 \\
 & 3 & 0.15148   & -1.20216  & 0.09695  & -0.31452 \\
1& 1 & 0.58587   & -0.19532  & 0.58336  & -0.19653  \\
 & 2 & 0.52890   & -0.61252  & 0.51960  & -0.59418  \\
 & 3 & 0.45908   & -1.08027  & 0.44000  & -0.93961  \\
2& 1 & 0.96729   & -0.19352  & 0.96669  & -0.19371  \\
 & 2 & 0.92716   & -0.59121  & 0.92504  & -0.58871  \\
 & 3 & 0.86109   & -1.01712  & 0.85890  & -0.97714
\end{tabular}
\end{ruledtabular}
\footnotetext[1]{from Ref.~\onlinecite{AnderssonQN}.}
\end{table}
He  translates this requirement into an infinite-fraction equation
involving the coefficients
 $\alpha$, $\beta$ and $\gamma$.
Majumdar and Panchapakesan gave an alternative condition using the
Hill determinant method \cite{mp}: a set of parameters $\{
l,\omega\}$ that give a convergent sum solve
\begin{equation}
D =
  \left|
 \begin{array}{ccccc} \beta_0 & \alpha_0 & \cdot & \cdot & \cdots
\\
  \gamma_1&\beta_1 & \alpha_1 & \cdot& \cdots  \\
  \cdot&\gamma_2 &\beta_2 & \alpha_2 &\cdots  \\
  \vdots&\vdots&\ddots &\ddots&\ddots \\
  \end{array}
 \right| =0.
\label{hill}
\end{equation}
This technique has the advantage of being more numerically
tractable than Leaver's, but it might be only valid for the Regge
poles that lie close to the real axis of the CAM plane. Both
Leaver's original method and the Hill determinant method have been
applied to the search for Regge poles in our study. We found
excellent agreement between Leaver's method and the
Hill-determinant method for small values of Regge mode index $n$.
We also agreed with Andersson's values given in
Ref.~\cite{Andersson2}. Fig.~\ref{fig:suivi-s0} exhibits the Regge
trajectories numerically calculated while Table~\ref{tab:table1}
presents a sample of QNM frequencies calculated from the
semiclassical formulas (\ref{sc1}) and (\ref{sc2}). A comparison
between the ``exact" and the semiclassical spectra shows a good
agreement, except for very low frequencies. Furthermore, the
semiclassical theory permits us to classify the resonances in
distinct families, each family being associated with one Regge
pole and therefore to understand the meaning of the indices $n$
and $\ell$ introduced to denote the QNM frequencies $\omega_{\ell
n}$.

At large $\omega$, the position of the Regge poles very closely
adheres to the asymptotic form:
\begin{equation}\label{asympRP}
\lambda_n (\omega) \approx 3\sqrt{3}M \omega + i\left(n-1/2
\right) \qquad n=1,2,3,\dots
\end{equation}
Eq.~(\ref{asympRP}) leads us to conclude that the $n$-th surface
wave is localized near the unstable circular photon orbit at
$R=3M$ by the following consistency argument: by reinstating
dependence on Schwarzschild time $t$ into Eq.~(\ref{ampliPR2}), we
can see that the surface waves circle the black hole in time $T =
2\pi \mathrm{Re} \ \lambda_n(\omega)/\omega \approx 2\pi 3
\sqrt{3} M$ for large $\omega$. Furthermore, a photon on the
circular orbit with constant radius $R$ takes the time $T'= 2\pi R
/ (1-2M/R)^{1/2}$ to circle the black hole (this result can be
found by integrating the Schwarzschild metric $ds^2 = 0$). By
equating $T$  and  $T'$, we obtain $R=3M$. Moreover, by using
Eqs.~(\ref{sc1}), (\ref{sc2}) and (\ref{asympRP}), we recover the
well-known high-frequency behaviors (see
Refs.~\onlinecite{Kokkotas} and~\onlinecite{Frolov-Novikov} and
references therein)
\begin{equation}\label{asympQNM}
\omega ^{(o)}_{\ell n} \approx \frac{\ell +1/2}{3\sqrt{3}M} \quad
\mathrm{and} \quad \frac{\Gamma _{\ell n}}{2} \approx
\frac{n-1/2}{3\sqrt{3}M}.
\end{equation}
These asymptotic behaviors have been obtained under the hypothesis
$\mathrm{Re} \ \lambda_n (\omega ) \gg \mathrm{Im} \ \lambda_n
(\omega )$ and are therefore valid for $\ell \gg n$.

\begin{figure}
\includegraphics[height=6cm,width=8cm]{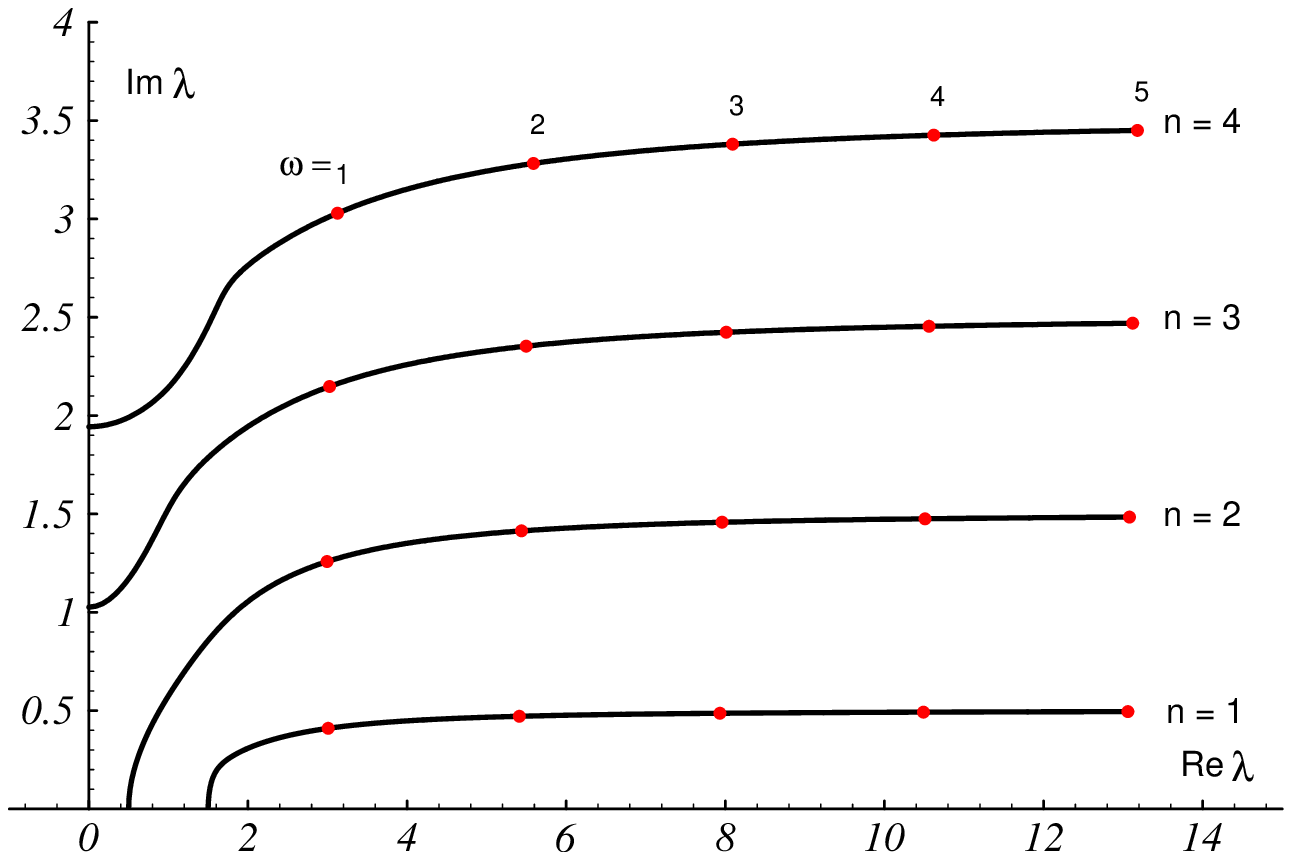}
 \caption{\label{fig:suivi-s2}The Regge poles $\lambda_n(\omega)$ followed
for $\omega=0\to 5$, $n=1,2,3,4$, for gravitational perturbations
($2M=1$). }
\end{figure}

Our approach also applies to electromagnetic and gravitational
scattering. In Fig.~\ref{fig:suivi-s2} and Table~\ref{tab:table2}
we present some results for the latter case. A comparison between
the exact and the semiclassical spectra shows a rather good
agreement.

Finally, it seems to us necessary to emphasize that the CAM method
we have developed here is based on two assumptions about the Regge
poles $\lambda_n (\omega )$: They must formally satisfy
$|\lambda_n (\omega )| \to +\infty $ as well as $\mathrm{Re} \
\lambda_n (\omega ) \gg \mathrm{Im} \ \lambda_n (\omega )$ as
$\omega \to \infty$.  As a consequence, our method is a
``high-frequency" approach which permits us to recover and to
physically interpret the spectrum of the QNM frequencies lying in
the region $2M \ \mathrm{Re} \ \omega
> 0.2$ of the complex $\omega$-plane. It should be noted that the
highly damped QNM (for gravitational perturbations) as well as
their associated frequencies which are connected with the area
spectrum of the black hole (see \cite{Hod98} for the first paper
on this subject and \cite{BaezNature} and references therein for
recent works in this domain) can neither be understood in the
framework of our approach nor interpreted in terms of surface
waves orbiting around the black hole near the unstable circular
orbit at $r=3M$: Indeed, when the index $n$ is very large, the
frequencies for these highly damped QNM are asymptotically given
by \cite{LeaverI, Nollert93, Motl2002, MotlNeitzke}
\begin{equation}\label{asympHDQNM}
\omega ^{(o)}_{\ell n} \approx \frac{ \ln 3}{8 \pi M} \quad
\mathrm{and} \quad \frac{\Gamma _{\ell n}}{2} \approx
\frac{n+1/2}{4M}
\end{equation}
and they therefore lie in the region $2M \ \mathrm{Re} \ \omega
\ll 0.2$ and $2M \ |\mathrm{Im} \ \omega| \gg 1$ of the complex
$\omega$-plane.

\section{Conclusion}

To conclude, we would like first to comment on some aspects of our
work and then to consider some possible extensions of the CAM
approach in gravitational physics:

- We have established the connection between Andersson's surface
waves and the QNM complex frequencies of the Schwarzschild black
hole in the particular context of wave scattering. This connection
is more general and could be also obtained directly from the
Regge-Wheeler equation, by extending the CAM method developed by
Sommerfeld \cite{Sommerfeld49} as an alternative to the Watson
approach \cite{Watson18}.

- Because the gravitational radiation created in many black hole
processes is dominated at intermediate timescales by QNMs, it can
be always interpreted in terms of surface waves. Whatever the
perturbation which modifies the geometry of a Schwarzschild black
hole, it leads to the excitation of surface waves localized close
to the unstable photon orbit. During its repeated
circumnavigations, a given surface wave decays by radiating away
its energy, leading to the damped ringing of the geometry. This
mechanism is analogous to the damped ringing of the Earth which
occurs during several days after a large earthquake and which can
be explained in term of the Rayleigh surface wave.
\begin{table}
\caption{\label{tab:table2} A sample of QNM frequencies for
gravitational perturbations ($2M=1$).}
\begin{ruledtabular}
\begin{tabular}{cccccc}
&       & \quad Exact\footnotemark[1] \quad & \quad
Exact\footnotemark[1] \quad
&   Semiclassical  &  Semiclassical   \\
 $\ell $ & $n$ &$   \omega ^{(o)}_{\ell n}  $
 &  $  \Gamma _{\ell n}/2  $
 & $  \omega ^{(o)}_{\ell n}   $ &  $ \Gamma _{\ell n}/2  $  \\
\hline
2& 1 & 0.74734  &-0.17793 & 0.75812 &  -0.17644  \\
 & 2 & 0.69342  &-0.54783 & 0.78134 &  -0.49976  \\
 & 3 & 0.60211  &-0.95655 & 0.77683 &  -0.83126   \\
3& 1 & 1.19889  &-0.18541 & 1.20332 &  -0.18455  \\
 & 2 & 1.16529  &-0.56260 & 1.20089 &  -0.54409   \\
 & 3 & 1.10337  &-0.95819 & 1.18284 &  -0.89846  \\
4& 1 & 1.61836  &-0.18832 & 1.62052 &  -0.18792  \\
 & 2 & 1.59326  &-0.56866 & 1.61119 &  -0.56020  \\
 & 3 & 1.54542  &-0.95982 & 1.58849 &  -0.92907
\end{tabular}
\end{ruledtabular}
\footnotetext[1]{from Chapter 4 of
Ref.~\onlinecite{Frolov-Novikov} or Refs.~\onlinecite{Kokkotas}
and~\onlinecite{Nollert}.}
\end{table}

- The CAM method could be naturally introduced in many other areas
of the physics of relativistic stars and black holes. Such a
program could include in particular the following topics:

\noindent (i) The interpretation of the QNM of neutron stars and
of the Kerr black hole in terms of surface waves. For the Kerr
black hole, such an approach could easily explain the splitting of
the quasinormal frequencies due to rotation.

\noindent (ii) The analysis of Hawking radiation as well as of
Kerr black hole superradiance.

\noindent (iii) The study of black holes immersed in
asymptotically anti-de Sitter space-times. The CAM approach could
provide new tests of the AdS/CFT correspondence recently proposed
in the context of superstring theory \cite{Maldacena}. With this
aim in view, the (2+1)-dimensional BTZ black hole \cite{btz} seems
to us very interesting because, in that particular space-time, the
wave equation can be solved exactly (see, e.g., \cite{Cardoso})
and therefore Regge poles can be analytically obtained.

\noindent (iv) The study of artificial black holes \cite{Novello}.
Here, it would be possible to benefit from the formidable CAM
machinery developed in electromagnetism, optics and acoustics.

\begin{acknowledgments}
It is a pleasure to acknowledge helpful discussions with Nils
Andersson.
\end{acknowledgments}

\newpage

\bibliography{RPandBH2}

\end{document}